\documentclass[twocolumn]{aastex63}
\usepackage{amssymb,amsmath}
\usepackage{graphicx}
\usepackage{xcolor,natbib}
\usepackage{multirow}
\usepackage[T1]{fontenc}
\usepackage{ae,aecompl}
\usepackage{newtxtext, newtxmath}
\usepackage{float}
\usepackage{subfigure}
\usepackage{longtable}
\usepackage{enumitem}
\usepackage{longtable,listings}
\usepackage[flushleft]{threeparttable}
\usepackage{parcolumns}
\usepackage{hyperref}

\def\obj{PG 1411+442}
\def\dif{\mathop{}\!\mathrm{d}}

\submitjournal{ApJ}
\shorttitle{550days Optical QPOs in PG1411+442}
\shortauthors{Zhang X. G.}

\begin{document}

\title{Optical QPOs with 550 day periodicity in the reverberation mapped broad line quasar PG 1411+442}

\correspondingauthor{Xue-Guang Zhang}
\email{xgzhang@gxu.edu.cn}

\author{Xue-Guang Zhang}
\affiliation{Guangxi Key Laboratory for Relativistic Astrophysics, School of Physical Science and Technology,
    GuangXi University, Nanning, 530004, P. R. China}

\begin{abstract} 
	In this manuscript, optical quasi-periodic oscillations (QPOs) with 550 day periodicity related to a 
candidate of sub-pc binary black hole (BBH) system are reported in the reverberation mapped broad line quasar 
PG 1411+442 but with different line profile of broad H$\alpha$ from that of broad H$\beta$ in its rms 
spectrum. First, considering sine function to describe the 18.8years-long light curves from the CSS, ASAS-SN and ZTF, 
550days periodicity can be confirmed with confidence level higher than 5$\sigma$. Second, the stable 550days optical QPOs 
can be re-confirmed with confidence levels higher than 5$\sigma$ by the Generalized Lomb-Scargle periodogram, the sine-like 
phase-folded light curves and the WWZ technique determined power maps. Third, based on simulated light curves by CAR process, 
confidence level higher than $3.5\sigma$ can be confirmed for the optical QPOs not related to intrinsic AGN variability. 
Moreover, considering spatial separation of central two BH accreting systems smaller than expected sizes of broad emission 
line regions (BLRs), central total BH mass higher than $10^6{\rm M_\odot}$ could lead to few effects of supposed
BBH systems on estimated virial BH masses. Meanwhile, disk precession is not preferred due to the similar estimated sizes 
of optical and NUV emission regions, and jet precession can be ruled out due to PG 1411+442 as a radio quiet quasar. The results 
strongly indicate it would be practicable by applying very different line profiles of broad Balmer emission lines to detect 
candidates of BBH systems in normal broad line AGN in the near future.  
\end{abstract}


\section{Introduction}


	Sub-pc binary black hole (BBH) systems are natural products in central regions of galaxies after considering galaxy 
merging as an essential process of hierarchical galaxy formation and evolution, as well discussed in \citet{bb80, sr98, mk10, 
fg19, mj22, ws23, ag24}. In order to identify candidates of sub-pc BBH systems, the long-standing optical 
Quasi-Periodic Oscillations (QPOs) with periodicities around hundreds to thousands of days are accepted as one of the 
preferred indicators. Here, some of the known optical QPOs related to candidates of sub-pc BBH systems are listed. 
\citet{gd15a, lg18, kp19} have reported reliable 1800days optical QPOs in the known quasar PG 1302-102, through 
its combined light curves covering more than 4 cycles. \citet{lg15} have reported 540days optical QPOs in PSO J334.2028+01.4075, 
through its combined light curves covering more than 5 cycles. \citet{gd15, cb16} have reported two samples of 
more than 160 optical QPOs with different periodicities, through the light curves covering at least 1.5 cycles. 
\citet{zb16} have reported 1500days optical QPOs in SDSS J0159, through its light curve covering around 2 cycles. 
\citet{ss20} have reported 1150days optical QPOs in Mrk915, through its light curve covering 3 cycles. \citet{ky20} 
have reported 1.2yr optical QPOs in Mrk231, through its light curve covering more than 10 cycles. \citet{lw21} 
have reported 1607days optical QPOs in SDSS J0252, through its combined light curve covering more than 4 cycles. 
\citet{zh22a, zh22c, zh23a} have reported 6.4yr optical QPOs in SDSS J0752 and 3.8yr optical QPOs in SDSS J1321 and 340days 
optical QPOs in SDSS J1609, through their light curves covering more than 2 cycles, 4 cycles and 4.5 cycles, 
respectively.

	Besides optical QPOs, different line profiles of broad Balmer emission lines can also be applied to identify 
candidates for sub-pc BBH systems in broad line AGN. We \citet{zh21d} have firstly shown the very different line profiles 
of broad Balmer lines accepted as an indicator of a central BBH system in SDSS J1547 with double-peaked broad H$\beta$ 
but single-peaked broad H$\alpha$. Meanwhile, 2159days optical QPOs can be confirmed in SDSS J1547 as shown in \citet{zh21d}. 
More recently, we \citet{zh23b} have shown that different line profiles of broad Balmer emission lines can be expected 
to have very different line width ratios from standard values, considering different effects of orbital obscuration on 
the two independent BLRs (broad emission line regions) related to the central sub-pc BBH systems in normal broad line 
AGN. And in \citet{zh23b}, we have shown the 1000days optical QPOs in the broad line AGN SDSS J1257 with the line width 
ratio 0.69 of broad H$\beta$ to broad H$\alpha$ very different from the standard value 1.1 for normal quasars as discussed 
in \citet{gh05}. Applications of different line profiles of broad Balmer emission lines to identify candidates 
for BBH systems can avoid effects of short time durations on reliability of optical QPOs, if only though optical QPOs to 
identify BBH systems.

\begin{figure*}
\centering\includegraphics[width = 18cm,height=10cm]{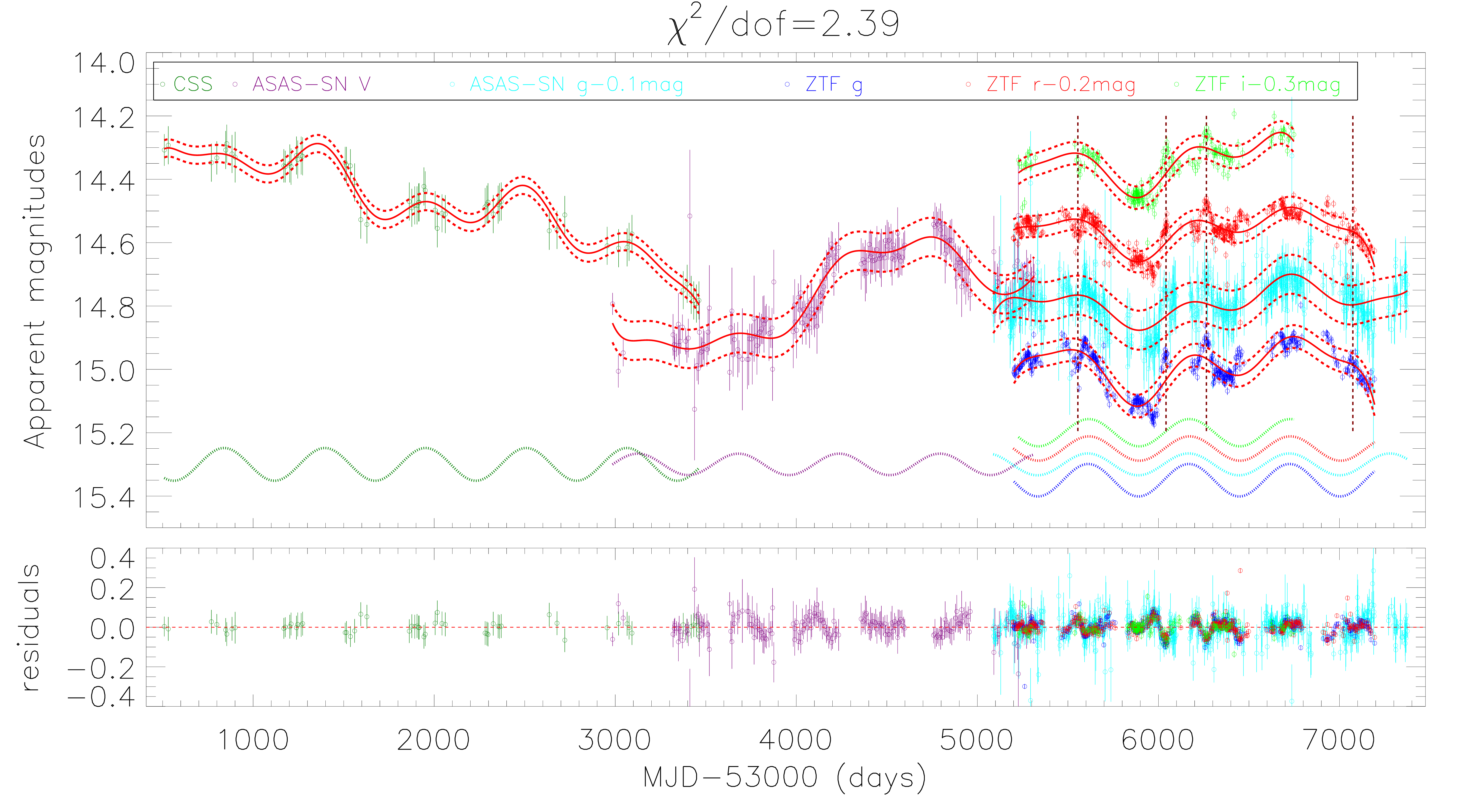}
\caption{Top panel shows the 18.8years-long light curves (open circles plus error bars) and the best descriptions (solid 
red lines) and the corresponding 1RMS scatters (dashed red lines) by the sine function plus trends described by 8th-degree 
polynomial functions. As shown in the legend, open circles in dark green, in purple, in cyan, in blue, in red and in green 
show the collected 1day binned data points from the CSS, the ASAS-SN V-band, the ASAS-SN g-band (minus 0.1mag), the ZTF 
g-band, the ZTF r-band (minus 0.2mag) and the ZTF i-band (minus 0.3mag), respectively. The dotted line in different color  
in bottom region of the top panel shows the determined sine component in the light curve shown as the same color. 
The vertical dashed lines in dark red mark four apparently sharp features in the light curves. Bottom panel shows the 
residuals calculated by the light curves minus the best descriptions, with horizontal dashed red line as the residuals 
equal to zero.}
\label{lmc}
\end{figure*}

\begin{figure*}
\centering\includegraphics[width = 18cm,height=6cm]{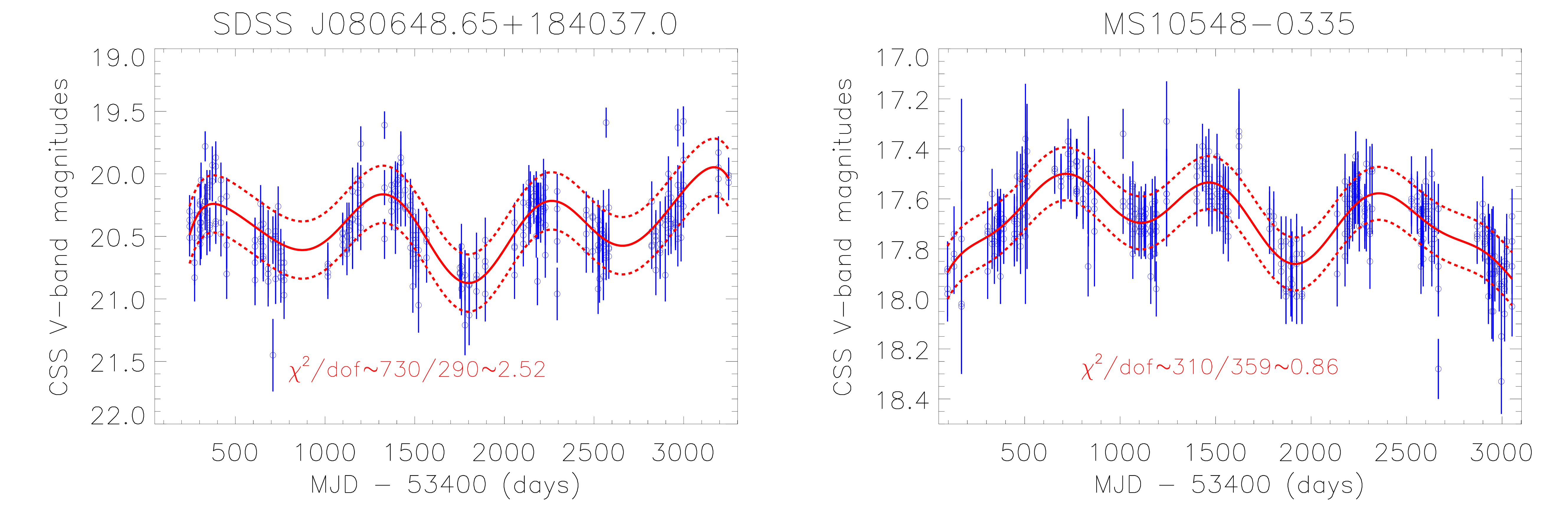}
\caption{Best fitting results and the corresponding $\chi^2/dof$ to the light curves of SDSS J080648.65+184037.0 
and MS 10548-0335 with reported QPOs in \citet{gd15}. In each panel, open circles plus error bars in blue show the light curve 
from the CSS, solid red line and dashed red lines show the best descriptions and the corresponding 1RMS scatters by the sine 
function plus a 8th-degree polynomial trend.}
\label{gra}
\end{figure*}

\begin{figure}
\centering\includegraphics[width = 8cm,height=5cm]{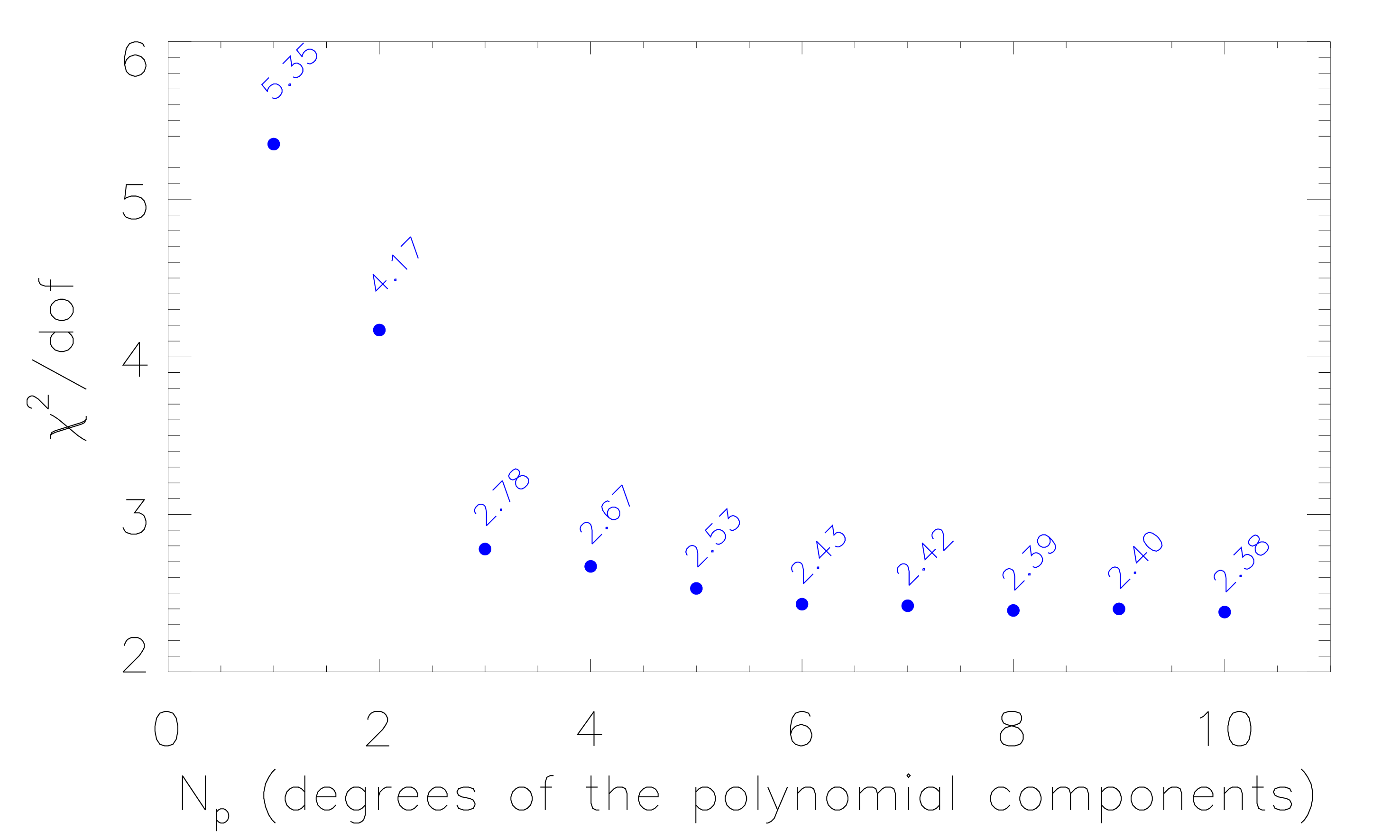}
\caption{On the dependence of $\chi^2/dof$ on different degrees for the polynomial components. Clear 
information of $\chi^2/dof$ has been marked around each data point.}
\label{np}
\end{figure}

	In the current stage, it is our research project to detect and identify more candidates for BBH systems 
in normal broad line AGN but with different line profiles of broad Balmer emission lines. Therefore, we start our research 
work among the known reverberation mapped broad line AGN with public spectroscopic results on broad emission lines. 
Among the reverberation mapped broad line AGN in \citet{pe04}, considering the measured line widths and the 
corresponding uncertainties, PG 1411+442 has unique properties of the varying components in broad Balmer emission lines 
in its rms spectrum (removing effects of none-variability components). Its broad H$\alpha$ has second moment $2437\pm196$km/s 
to be about $1.52_{-0.25}^{+0.32}$ times of second moment $1607\pm169$km/s of broad H$\beta$, however the broad H$\alpha$ 
has full width at half maximum (FWHM) $1877\pm375$km/s to be about $0.78_{-0.23}^{+0.32}$ times of FWHM $2398\pm353$km/s of 
the broad H$\beta$. The very different line width ratios strongly indicate very different line profile of the varying component 
in the broad H$\alpha$ from that in the broad H$\beta$ in the rms spectrum of PG 1411+442. Therefore, it is interesting to 
check long-term variability properties of PG 1411+442, to determine expected optical QPOs, motivated by its very different 
line profiles of the varying components in the broad Balmer emission lines in its rms spectrum, which is the main objective 
of this manuscript. Section 2 presents main results on the long-term optical variability of PG 1411+442. Section 3 shows 
the necessary discussions. Section 4 gives final summary and main conclusions. In this manuscript, the cosmological parameters 
have been adopted as $H_{0}=70{\rm km\cdot s}^{-1}{\rm Mpc}^{-1}$, $\Omega_{\Lambda}=0.7$ and 
$\Omega_{\rm m}=0.3$.

\begin{figure}
\centering\includegraphics[width = 8cm,height=15cm]{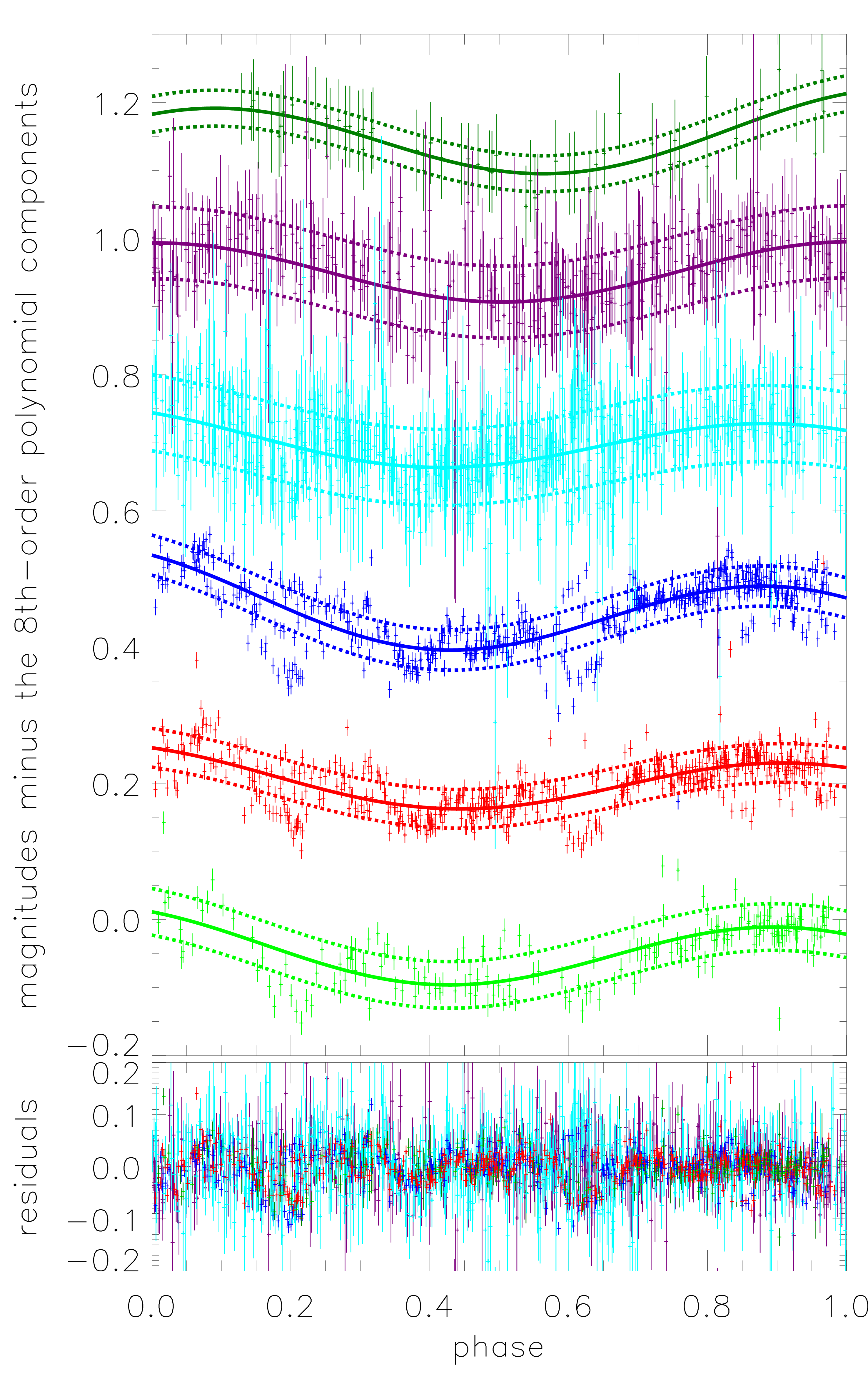}
\caption{Top panel shows the phase folded light curves (crosses plus error bars in different colors) after removing the 
polynomial trends and the best descriptions (solid lines in different colors) and the corresponding 1RMS scatters (dashed 
lines in different colors) by the sine function. In top panel, symbols and line styles in dark green, in purple, in cyan, 
in blue, in red and in green show the corresponding results for the light curves from the CSS (plus 1.15mag), ASAS-SN V-band 
(plus 0.95mag), ASAS-SN g-band (plus 0.7mag), and ZTF g/r/i-band (plus 0.45mag, plus 0.2mag, minus 0.05mag), respectively. 
Bottom panel shows the corresponding residuals (in the same color) calculated by the folded light curve minus the best 
descriptions.}
\label{l2mc} 
\end{figure}

\section{Optical QPOs in PG 1411+442}

        PG 1411+442 at $z\sim0.0896$ is a known reverberation mapped broad line AGN discussed in \citet{kas00, pe04}. The 
18.8years-long light curves of PG 1411+442\ shown in Fig.~\ref{lmc} can be collected from the Catalina Sky Survey (CSS) 
\citep{dr09} with MJD-53000 from 509.35 (May, 2005) to 3462.28 (June, 2013), from the All-Sky Automated Survey for Supernovae 
(ASAS-SN) \citep{sp14, ks17} with MJD-53000 from 2985.10 (March, 2013) to 7374.81 (March, 2024), from the Zwicky Transient 
Facility (ZTF) \citep{bk19, ds20} with MJD-53000 from 5202.37 (March, 2018) to 7194.16 (September, 2023). Then, similar as 
what we have recently done in \citet{zh23a}, the following four methods are applied to check whether are there optical QPOs 
in PG 1411+442.

	For the first method (the direct-fitting method), a 8th-degree polynomial function plus a sine component with 
periodicity $T_q$ are applied to describe each light curve $LC_{k,t}$ from the different sky survey projects,
\begin{equation}
LC_{k,t}~=~\sum_{i=0}^{8}(a_{k,i}\times t_{k}^i)~+~g_k\times\sin(\frac{2\pi~t_{k}}{T_q}~+~\phi_{0,k})
\end{equation},
with the suffix $k$ meaning which light curve (CSS V-band, ASAS-SN V-band, ASAS-SN g-band, ZTF g-band, ZTF r-band, ZTF i-band) 
is used, and the parameters $a_{k,i}$ as the polynomial coefficients. When the six light curves are being simultaneously 
described by the model functions above, the parameters $a_{k,i}$, $g_k$ and $\phi_{0,k}$ are different but $T_q$ are fixed 
for the light curves from the different sky survey projects. Then, through the Levenberg-Marquardt least-squares minimization 
technique (the known MPFIT package) \citep{mc09}, the best descriptions and the corresponding residuals (the light curves 
minus the best descriptions) can be determined and shown in Fig.~\ref{lmc} with $\chi^2/dof\sim2.39$ ($dof$ as the degree of 
freedom). The determined periodicity and the corresponding 1$\sigma$ uncertainty are $550\pm2$days. Although 
the $\chi^2/dof\sim2.39$ is larger than 1, the determined fitting results can be accepted, mainly due to the following main 
reason. The collected light curves are not smooth enough but with some sharp features (probably related to intrinsic AGN 
variability), therefore the applied polynomial component plus sine components cannot lead to the determined fitting results 
with $\chi^2/dof\sim1$. In top panel of Fig.~\ref{lmc}, four apparently sharp features in the light curves are marked by 
vertical dashed lines in dark red. Moreover, as a comparison, two light curves from CSS are collected for the SDSS 
J080648.65+184037.0 and MS 10548-0335 reported in the sample of \citet{gd15} to have robust QPOs with periodicities around 
890days. Based on a 8th-degree polynomial component plus a sine component, the best descriptions to the two light curves 
are shown in Fig.~\ref{gra}, with the determined $\chi^2/dof$ about 2.5 in SDSS J080648.65+184037.0 but about 0.86 in MS 
10548-0335. Therefore, although the determined fitting results lead $\chi^2/dof$ to be apparently larger than 1, the shown 
fitting results in Fig.~\ref{lmc} can be reasonably accepted.

	Here, in order to determine the degree of the applied polynomial components, different degrees $N_p$ 
have been applied, leading to the corresponding values $\chi^2/dof$. As shown in Fig.~\ref{np} for the dependence of 
$\chi^2/dof$ on $N_p$, there are none apparent variability of $\chi^2/dof$ for $N_p~\ge~8$. Therefore, the 8th degree 
polynomial components are accepted. Furthermore, besides the 8th-degree polynomial functions plus sine components, a 
30th-degree polynomial functions without considering any sine components applied to describe the light curves can lead 
to the $\chi^/dof\sim2.5$ related to the new model polynomial functions. Then, through the F-test technique 
as what we have recently done in \citet{zh23a}, the sine components are preferred with confidence levels higher than 
7$\sigma$, rather than pure applications of higher degree polynomial functions.

\begin{figure}
\centering\includegraphics[width = 8cm,height=16cm]{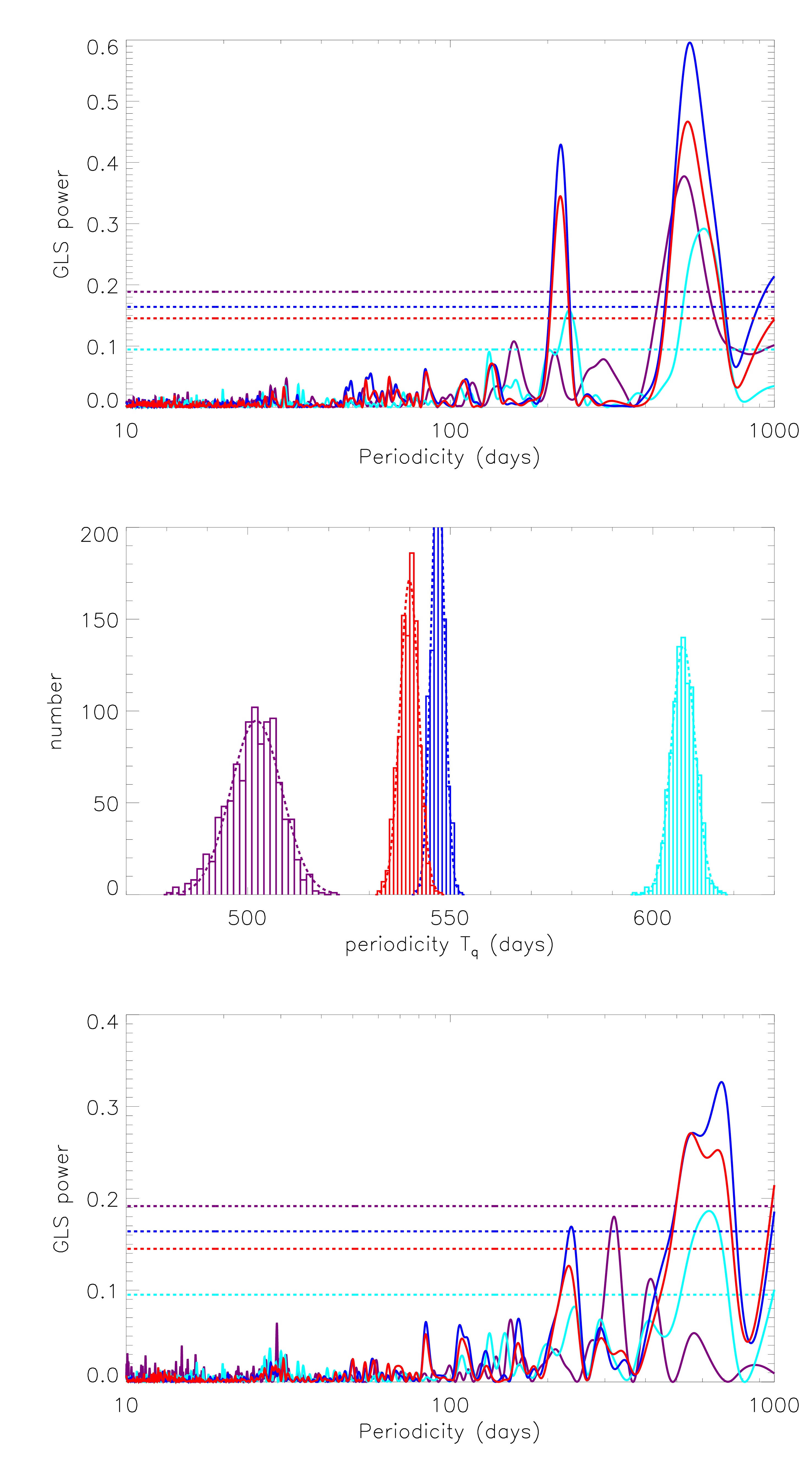}
\caption{Top panel shows the power properties through the Generalized Lomb-Scargle periodogram. In top panel, solid line 
in purple, in cyan, in blue and in red represent the results through the ASAS-SN V-band line curve, ASAS-SN g-band light 
curve, ZTF g/r-band light curves, after removing the polynomial trends. The corresponding $5\sigma$ confidence level is 
shown as the horizontal dashed line in the same color. Middle panel shows the distributions of GLS determined periodicities 
through the bootstrap method. In middle panel, histogram in purple, in cyan, in blue and in red show the corresponding results 
through the ASAS-SN V-band line curve, ASAS-SN g-band light curve, ZTF g/r-band light curves. And the thick dashed line in 
the same color shows the Gaussian described results to the corresponding distribution. Bottom panel shows the 
power properties through the light curves without subtractions of any polynomial trends. In bottom panel, line styles have 
the same meanings as those in the top panel.}
\label{gls}
\end{figure}

\begin{figure}
\centering\includegraphics[width = 8cm,height=15cm]{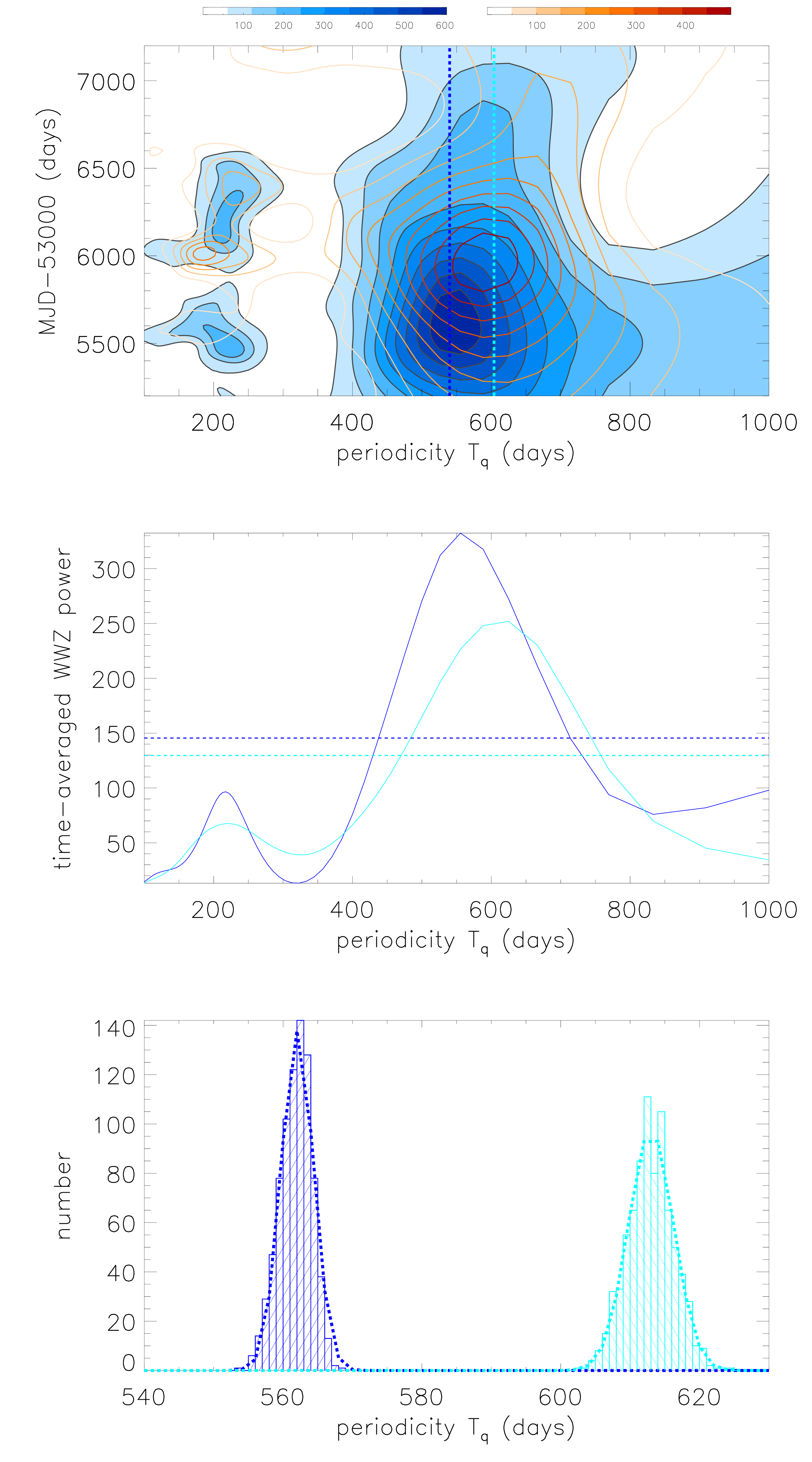}
\caption{Top panel shows the two dimensional power maps determined by the WWZ method. Middle panel shows the WWZ method 
determined time-averaged power properties. Bottom panel shows the bootstrap method determined periodicity distributions. 
In top panel, the vertical dashed lines mark the WWZ method determined periodicities, contour filled with bluish colors 
represent the results through the ZTF g-band light curve after removing the polynomial trend, contour with levels shown 
in reddish colors represent the results through the ASAS-SN g-band light curve after removing the polynomial trend. In 
top regions of top panel, color bars show the corresponding number densities for the contour levels. In middle panel, the 
horizontal dashed blue line and horizontal dashed purple line mark the corresponding 5$\sigma$ confidence levels for the 
periodicities through the Monte Carlo method. In bottom panel, histogram filled by blue lines and filled by purple lines 
show the WWZ determined periodicity distributions by the bootstrap, through the ZTF g-band and ASAS-SN g-band light cures 
after removing the polynomial trends, respectively, and thick dashed line in the same color shows the Gaussian described 
results to the corresponding distribution.}
\label{wwz}
\end{figure}

	For the second method, after subtracting the trends described by the 8th-degree polynomial functions, the corresponding 
phase-folded light curves with accepted the determined periodicity $550\pm40$days are shown in Fig.~\ref{l2mc}, which can be 
described by the sine function $\sin(2\pi~t+\phi_0)$. As an extension of the directing fitting method, application 
of the phase-folded method can show QPOs more intuitively in phase space, to support the optical QPOs in the light curves 
of PG 1411+442.

	For the third method, the widely accepted Generalized Lomb-Scargle (GLS) periodogram \citep{ln76, sj82, zk09, vj18} 
(python package of astroML.time\_series, \url{https://www.astroml.org/_modules/astroML/time_series/periodogram.html#lomb_scargle}) 
is applied to the light curves 
after subtractions of the polynomial trends. Here, due to small number of data points in the CSS light curve and short time 
duration of the ZTF i-band light curve, the GLS periodogram is not applied to the CSS light curve nor to the ZTF i-band 
light curve. Top panel of Fig.~\ref{gls} shows the GLS power properties. It is clear for the periodicity around 500-600days 
with confidence level higher than $5\sigma$ (the false-alarm probability of 5e-7) determined by the bootstrap method as 
discussed in \citet{ic19}. Meanwhile, as shown in the middle panel of Fig.~\ref{gls}, the well-known bootstrap method within 
1000 loops is applied to determine the periodicities and the corresponding uncertainties about 502$\pm7$days, 607$\pm$3days, 
547$\pm$2days and 540$\pm2$days for the ASAS-SN V-band, the ASAS-SN g-band and the ZTF g/r-band light curves, respectively. 

	Before proceeding further, two additional points are noted. First, through the collected light curves from different 
sky survey projects, the determined periodicities by the GLS periodogram (also by the following WWZ technique) are a bit 
different, probably due to effects of intrinsic AGN variability. More detailed discussions on the effects of AGN variability 
on the determined periodicities related to sub-pc BBHs systems will be given in one manuscript which we have submitted to ApJ. 
Second, the GLS periodogram is also applied to the collected ZTF g/r-band light curves and ASAS-SN V/g-band light curve, 
without subtractions of any polynomial trends from the light curves. The results can be applied to confirm that the 
subtractions of polynomial trends have few effects on our determined periodicities but can lead to more apparent signs for 
QPOs. The corresponding GLS power properties are shown in the bottom panel of Fig.~\ref{gls}. It is clear that apparent peaks 
around 500-600days with confidence level higher than 5$\sigma$ can be confirmed in the GLS power properties through the ZTF 
g/r-band and ASAS-SN g-band light curves. However, there is no reliable periodicity with confidence level higher than 5$\sigma$ 
in the GLS power properties through the ASAS-SN V-band light curve. Therefore, the applied polynomial trends lead to the same 
but more significant peaks in the GLS power properties through the light curves.

	For the fourth method, the WWZ (Weighted wavelet z-transform) technique \citep{fg96, al16, gt18, ks20, ly21} (
the python code wwz.py written by M. Emre Aydin, \url{https://github.com/eaydin/WWZ/blob/master/wwz.py}) is applied to check 
the optical periodicities in PG 1411+442 with frequency step of 0.0001 and searching periodicities from 100 days to 1000 days. 
The corresponding results are shown in Fig.~\ref{wwz} with $5\sigma$ confidence levels determined by common Monte Carlo method 
applied through 3.2 million randomly created light curves by white noise process. Here, due to small number of data points in 
the CSS light curve and large time gaps in the ASAS-V band light curve and short time duration of the ZTF i-band light curve, 
the WWZ technique is not applied to the CSS, ASAS-V band and ZTF i-band light curves. Meanwhile, due to totally similar WWZ 
determined results for the ZTF g-band light curve and the ZTF r-band light curve, only the WWZ determined results for the ZTF 
g-band light curve are shown in Fig.~\ref{wwz}. Clearly similar periodicities around 562$\pm$3days in the ZTF g/r-band light 
curves and 613$\pm$3days in the ASAS-SN g-band light curve can be confirmed as shown in the bottom panel of Fig.~\ref{wwz} 
through the bootstrap method.

	Therefore, the optical QPOs with periodicities around 500-600days (mean value and uncertainty as $\sim$550$\pm$50days) 
in PG 1411+442 can be detected from the 18.8years-long light curves (time duration about 12.5 times longer than the detected 
periodicities) with confidence level higher than $5\sigma$, based on the best-fitting results directly by the sine function 
shown in Fig.~\ref{lmc}, on the sine-like phase-folded light curve shown in Fig.~\ref{l2mc}, on the results of GLS periodogram 
shown in Fig.~\ref{gls}, and on the results determined by the WWZ technique shown in Fig.~\ref{wwz}.

	Furthermore, the similar procedure as recently done in \citet{zh23a} and also similar as done in \citet{vu16} is 
applied to test whether the optical QPOs were actually related to intrinsic AGN variability of PG 1411+442. Based 
on the CAR (the first order Continuous AutoRegressive) process in \citet{kbs09}, 
\begin{equation}
\dif LC_{t}=\frac{-1}{\tau}LC_{t}\dif t+\sigma_{*}\sqrt{\dif t}\epsilon(t)~+~LC_{0}
\end{equation}
with $\epsilon(t)$ as a white noise process with zero mean and variance equal to 1, 16660 ($1/(1-P_4)$ with $P_4$ as the 
$4\sigma$ probability) light curves $LC_{t}$ with the same time information of CSS, ASAS-SN V-band and ASAS-SN g-band light 
curves as shown in Fig.~\ref{lmc} are created, with variability timescale $\tau/days$ randomly selected from 50days to 
1000days and variability amplitude $\sigma_*/(mag/day^{0.5})$ randomly selected from 0.006mag/day$^{0.5}$ to 0.03mag/day$^{0.5}$, 
with the timescales and amplitudes being the common values for normal quasars as shown in \citet{kbs09, koz10, mi10} and with 
$LC_{0}=14.83$ (the mean magnitude of the combined CSS, ASAS-SN V-band and ASAS-SN g-band light curves). Then, among the 16660 
light curves, there are only 8 light curves with probably QPOs with periodicities around 550$\pm$100days, detected by the GLS 
method with confidence levels higher than 5$\sigma$ and determined by the Equation (1) leading to the corresponding 
$\chi^2/dof~<~3$. Therefore, the probability is lower than $4.8\times10^{-4}$ (8/16660) to detect fake QPOs through intrinsic 
AGN variability. In other words, the confidence level higher than $3.5\sigma$ can be confirmed that the detected optical QPOs 
in PG 1411+442 are not related to intrinsic AGN variability.

	Here, the first order CAR process is applied above. However, as discussed in \citet{kb14, kv17, mv19, yr22, kg24}, 
higher order CARMA(p, q) ($p\ge1$ and $q\le p$) (Continuous AutoRegressive Moving Average) process rather than the simple first 
order CAR process (=CARMA(1,0)) should be possibly preferred to describe AGN variability. Meanwhile, the CARMA(p, q) process 
has been applied to test robustness of detected QPOs, such as the discussions in \citet{bg20, tz20}. Therefore, it is necessary 
to check whether higher order CARMA(p, q) is preferred than the applied first order CAR process in the PG 1411+442. Based on 
the linear interpretation applied to the high quality ZTF g-band light curve of PG 1411+442, an evenly sampled light curve can 
be created with the same number of data points as that of the ZTF g-band light curve. Then, through both the Bayesian 
Information Criterion (BIC) and the Akaike Information Criterion (AIC), the orders (p=1, q=0) of the CARMA(p, q) process for 
the evenly sampled light curve can be determined through applications of the function arma\_order\_select\_ic in python package 
statsmodels (\url{https://tedboy.github.io/statsmodels_doc/doc/generated/statsmodels.tsa.stattools.arma_order_select_ic.html}). 
Therefore, it is efficient enough for applications of the CAR process to trace intrinsic AGN variability in PG 1411+442.

	Before ending the section, two additional points should be noted. For the first point, as shown in Fig.~\ref{lmc}, 
there really are large time gaps in the CSS light curve, therefore only the direct-fitting method is applied to the 
CSS light. The best fitting results with the same periodicity not only to the CSS light curve but also to the light curves 
from the other sky survey projects actually provide more stable evidence to support the optical QPOs in PG 1411+422, through 
the direct-fitting method. For the second point, as shown in Fig.~\ref{l2mc}, there are not the same quality for the folded 
light curves. However, considering different intrinsic AGN variability properties in different epochs and also the different 
time gaps in different light curves, the results in Fig.~\ref{l2mc} can be reasonably accepted. 

\section{main discussions}

	We firstly discuss whether an expected BBH system has effects on estimations of virial BH mass of reverberation 
mapped broad line AGN, which will provide further clues on the conditions that broad emission line properties 
can be applied to estimate total BH mass of a broad line AGN even harboring a BBH system.

	Based on the optical QPOs with periodicity about 550days in PG 1411+442, assumed a BBH system as discussed in 
\citet{eb12}, the expected space separation of the central BBH system in PG 1411+442 can be estimated as
\begin{equation}
	S_{BBH}~\sim~0.432M_{8}(\frac{T_q/year}{2652M_{8}})^{2/3}~pc~\sim~5.8_{-2.8}^{+4.8}{\rm light-days}
\end{equation}
with $M_8$ as the BH mass in units of $10^8{\rm M_\odot}$ and with the accepted virial BH mass 
$(4.43\pm1.46)\times10^8{\rm M_\odot}$ reported in \citet{pe04}. If only the space separation is very smaller than BLRs 
(broad emission line regions) in the central region of PG 1411+442, the dynamical properties of broad emission lines can 
be efficiently applied to determine total virial mass of central binary system. Considering the sizes of BLRs about 70-120 
light-days of \obj~ in \citet{pe04} at least 10 times larger than the $S_{BBH}$ in PG 1411+442, therefore, the estimated 
virial BH mass is efficient enough, simply accepted the two BLRs have been at least partly mixed similar as discussed 
in \citet{sl10}.

\begin{figure*}
\centering\includegraphics[width = 18cm,height=12cm]{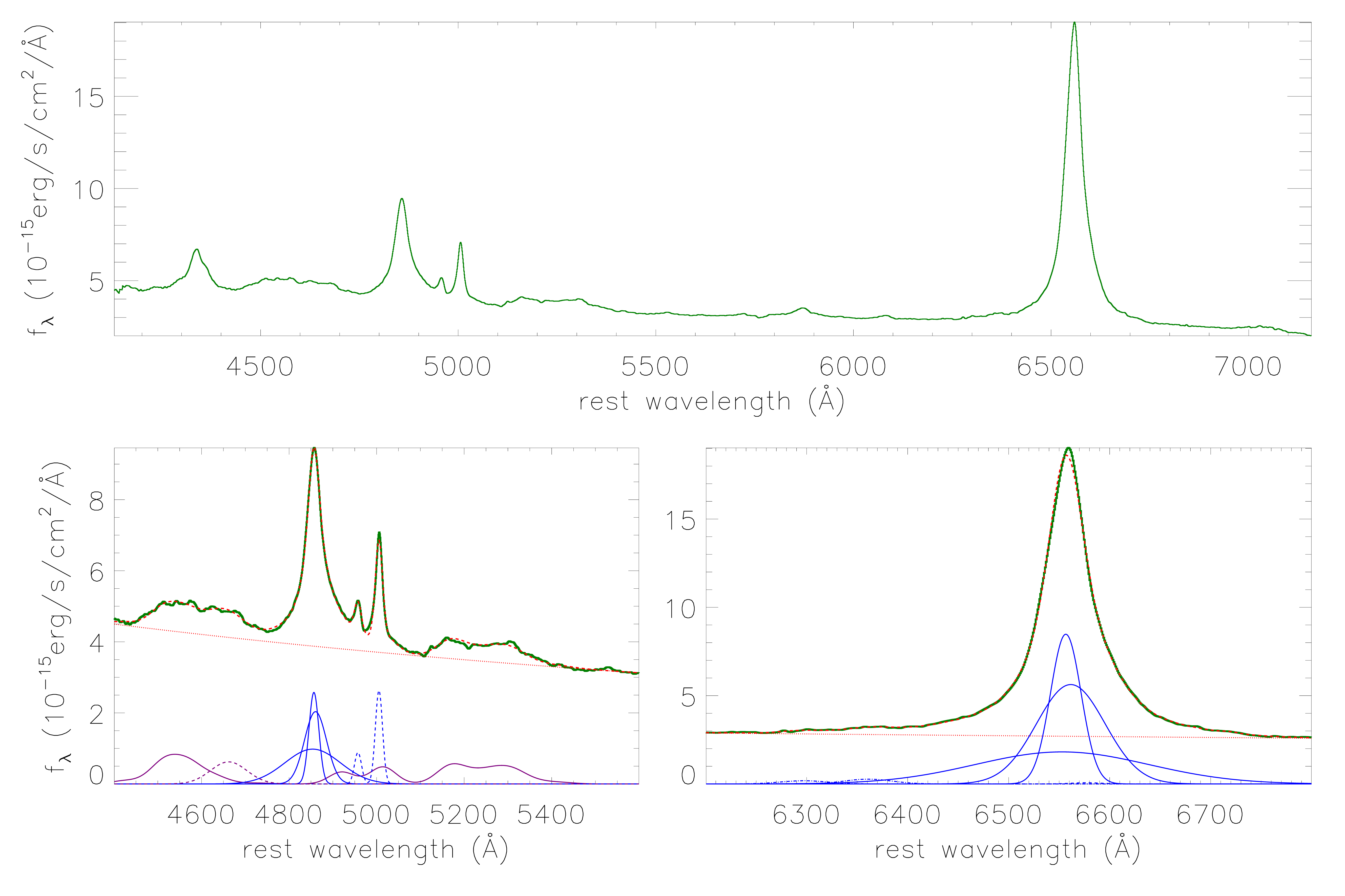}
\caption{Top panel shows the average spectrum of PG 1411+422 collected from \citet{kas00}. Bottom panels show the best fitting 
results (dashed red lines) to the emission lines (solid dark green lines) around H$\beta$ and around H$\alpha$. In each bottom 
panel, dotted red line shows the determined power law continuum emissions, solid blue lines show the determined Balmer emission 
lines including three Gaussian components. In bottom left panel, solid purple line shows the determined optical Fe~{\sc ii} 
emission lines, dashed purple line shows the determined broad He~{\sc ii} component, dashed blue lines show the determined 
[O~{\sc iii}] doublet.}
\label{spec}
\end{figure*}

	Furthermore, simply considering the dependence of BH mass on continuum luminosity reported in \citet{pe04} and the 
dependence of BLRs sizes $R_{BLRs}$ on continuum luminosity reported in \citet{kas00, bd13}, the following formula can be found 
\begin{equation}
\begin{split}
\log(M_{8})~&=~-0.12~+~0.79~\times~\log(\frac{L_{con}}{\rm 10^{44}}) \\
\log(\frac{R_{BLRs}}{\rm light-days})~&=~1.555~+~0.542~\times~\log(\frac{L_{con}}{\rm 10^{44}})
\end{split}
\end{equation}.
In order to find $S_{BBH}~\simeq~R_{BLRs}$, we will have 
\begin{equation}
	\log(M_{8})~\simeq~1.89\log(\frac{T_q}{\rm year}) - 3.43
\end{equation}.
Simply considered $M_{BH}\sim10^8,~10^7,~10^6{\rm M_\odot}$, we will have $T_q\sim$220years,~19years,~5.7years. In one word, 
unless central total BH masses were around $10^6{\rm M_\odot}$ or smaller leading to the expected $S_{BBH}$ larger than 
$R_{BLRs}$, central BBH systems have few effects on estimations of virial total masses of central BHs.

	We secondly discuss the other explanations to the optical QPOs in PG 1411+442, besides the assumed central BBH system. 

	The precessions of emission regions with probable hot spots for the optical emissions can also be applied to describe 
the detected optical QPOs in \obj. As discussed in \citet{eh95, st03}, the expected disk precession period can be estimated as 
\begin{equation}
T_{\rm pre}\sim1040M_{8}R_{3}^{2.5}yr
\end{equation},
with $R_{3}$ as distance of optical emission regions to central BH in units of 1000 Schwarzschild radii ($R_g$). Considering 
the optical periodicity about 550days and the BH mass about $(4.43\pm1.46)\times10^8{\rm M_\odot}$, the expected $R_3$ could 
be around 0.04\ in \obj. However, based on the discussed distance of NUV emission regions to central BHs in \citet{mc10} 
through the microlensing variability properties of eleven gravitationally lensed quasars, the NUV 2500\AA~ continuum emission 
regions in \obj~ have distance from central BH as
\begin{equation}
\log{\frac{R_{2500}}{cm}}=15.78+0.80\log(\frac{M_{BH}}{10^9M_\odot})
\end{equation}
leading size of NUV emission regions to be about $48R_g$. The estimated NUV emission regions have similar distances as the 
optical continuum emission regions in \obj~ under the disk precession assumption, indicating that the disk precessions of 
emission regions are not preferred in \obj.

	Meanwhile, long-standing QPOs can be detected in blazars due to jet precessions as discussed in \citet{sc18, bg19, 
oa20}. \obj~ is covered in Faint Images of the Radio Sky at Twenty-cm \citep{bw95, hw15} with peak flux about 1.61mJy, leading 
to the corresponding radio loudness about 0.01, indicating PG 1411+442 as a radio quiet object. Therefore, jet precessions 
can be well ruled out in \obj.

	We thirdly try to discuss spatial properties of the centra BBH system through spectroscopic emission line features in 
PG 1411+442. Unfortunately, through the optical average spectrum in \citet{kas00} and the single-epoch spectrum in 
\citet{bg92} of PG 1411+442, there are no unique features in broad Balmer lines probably related to the central BBH system, 
such as double-peaked features, very asymmetric features, etc., as the shown spectroscopic results in Fig.~\ref{spec}. Here, 
we do not show further discussions on our model functions including multiple Gaussian functions plus optical Fe~{\sc ii} 
template to describe the emission lines around H$\beta$ and around H$\alpha$ which is beyond the scope of the manuscript, 
detailed descriptions on our model functions can be found in our paper \citet{zh21b}. Therefore, there are no further 
discussions on the probable BBH system through spectroscopic emission features of PG 1411+442. Moreover, as discussed in 
\citet{bs08, sl10, dh15, ng20, ji21}, BBHs can lead to variability of broad emission line profiles. However, considering 
variability of central ionization continuum emissions leading to deeper/shallower ionization boundaries in the BLRs (to change 
geometric structure of BLRs for broad line emissions), variability of broad emission line profiles can be well expected, 
not due to central BBHs but due to common intrinsic AGN variability, such as the known variability of broad emission lines 
of the reverberation mapped broad line AGN. Therefore, at present stage, we do not show further discussions on variability of 
spectroscopic emission features of PG 1411+442 probably related to BBH system, due to loss of definite methods to determine 
effects of intrinsic AGN variability on variability of broad emission lines. However, the results in this manuscript provide 
clues to support indicators for BBHs by different line profiles in not only single-epoch spectrum but also in rms spectrum 
through multi-epoch spectra.

\section{Summary and Conclusions}

	Motivated by optical QPOs candidates expected in broad line AGN with very different line profiles of broad Balmer 
emission lines, the known reverberation mapped broad line AGN PG 1411+442 is selected as the subject of this manuscript, due 
to its very different dynamical properties of the varying components in the broad Balmer emission lines in its rms spectrum. 
Through the 18.8years-long light curves collected from the CSS, ASAS-SN and ZTF, stable 550days optical QPOs can be confirmed 
with confidence levels higher than 5$\sigma$ in PG 1411+442, through different methods/techniques. Furthermore, based on the 
CAR process, confidence level higher than 3.5$\sigma$ can be confirmed that the optical QPOs are not related to intrinsic AGN 
variability in PG 1411+442. The robust optical QPOs can be applied to support a BBH system candidate in the 
PG 1411+442. The results not only provide clues to support that even central BBH system candidates expected in 
normal broad line AGN, the virial technique can also be efficiently applied to estimate central total BH masses, but also 
provide confident clues for searching optical QPOs among broad line AGN with very different line profiles of broad Balmer 
emission lines in the near future.

\acknowledgments
Zhang gratefully acknowledge the anonymous referee for giving us constructive comments and suggestions to greatly 
improve our paper. Zhang gratefully acknowledges the kind grant support from NSFC-12173020 and NSFC-12373014. This paper 
has made use of the data from the ZTF \url{https://www.ztf.caltech.edu}, from ASAS-SN 
\url{https://www.astronomy.ohio-state.edu/asassn/index.shtml}, from CSS \url{https://catalina.lpl.arizona.edu/}. This 
research has made use of the NASA/IPAC Extragalactic Database (\url{http://ned.ipac.caltech.edu/classic/}) funded by the 
National Aeronautics and Space Administration and operated by the California Institute of Technology.


\end{document}